# Double-Sided Energy Auction: Equilibrium Under Price Anticipation

M.Nazif Faqiry and Sanjoy Das

Electrical & Computer Engineering Department
Kansas State University

*Abstract*—This paper investigates the problem of proportionally fair double-sided energy auction involving buying and selling agents. The grid is assumed to be operating under islanded mode. A distributed auction algorithm that can be implemented by an aggregator, as well as a possible approach by which the agents may approximate price anticipation is considered. Equilibrium conditions arising due to price anticipation is analyzed. A modified auction to mitigate the resulting loss in efficiency due to such behavior is suggested. This modified auction allows the aggregate social welfare of the agents to be arbitrarily close to that attainable with price taking agents. Next, equilibrium conditions when the aggregator collects a surcharge price per unit of energy traded is examined. A bi-objective optimization problem is identified that takes into account both the agents' social welfare as well as the aggregator's revenue from the surcharge. Results of extensive simulations, which corroborate the theoretical analysis, are reported.

## I. INTRODUCTION

NETWORK economics has made automated trading feasible, where resources are exchanged for money through transactions that are done entirely through software *agents* and without the need for any human intervention. It is often the case that the agents participating in the trade are involved in direct competition with one another, where the objective of each agent is to maximize its own *payoff*. In these situations the trade proceeds through an *auction* mechanism. In single-sided auctions all agents are either *buyers* that compete to acquire a finite resource, or *sellers* that compete to sell their goods. *Double auctions* are mechanisms involving both buyers and sellers, which simultaneously participate in the bidding process and are allocated individual shares of the resource.

Recent technological advancements in communications and renewable energy generation have created much research interest in energy auction algorithms [SA+15, PD+15]. In these mechanisms, the agents may represent individual domestic units within a microgrid, with energy representing the resource, and with PV-equipped homes experiencing a surplus of energy acting as sellers, and the remaining domestic units as buyers [MF+14]. An agent may also represent an individual microgrid involving a community of homes that collectively behave as a single unit in the ensuing auction [WH15]. A buyer agent's payoff is typically the difference between its *utility* gained from consuming a certain amount of energy and the price that it has to pay in order to procure that energy. Likewise a seller agent's payoff may be formulated as the sum of the monetary gain from supplying an amount of energy and utility it gains from retaining any surplus energy that is not traded.

The Kelly mechanism refers to a class of auction algorithms where agents are allowed to place individual bids on the resource, while a separate auctioneer that receives these bids, allocates the resource share of each bidding agent in proportion to the bid values [KMT98]. With a large number of agents, such a *proportional allocation* mechanism has been shown to maximize the aggregate utilities of all agents, the latter commonly referred to as the *social welfare* [MB03, JT04, CST15]. Such social welfare maximizing mechanisms are *efficient* auctions.

Unfortunately, the underlying assumption for proportional allocation to be efficient is that the agents be *price takers*, i.e. ones that assume that the bids that they place do not influence the market price of the resource. While this is approximately true for auctions involving a large number of agents, in smaller auctions, agents are aware of their own *market power*, and accordingly, place strategic bids on the resource. Such a *price anticipatory* bidding results in a *loss in efficiency* where the resource allocation of the auction no longer maximizes the social welfare.

It must be noted that there are other efficient auctions that explicitly focus on eliciting truthful bidding from the bidders; the most significant ones being based on the well-known Vickrey-Clarke-Groves mechanism [NR07]. Furthermore, auctions designed to maximize the auctioneer's own earned revenue have been proposed in the vast body of literature on this subject. However, these mechanisms are not of direct relevance in this research, which focuses on efficient double auctions along with proportional allocation of electricity among its buyers.

### A. Current Research

Proportional allocation in single-sided buyers' auctions has been rigorously analyzed [JT04, CSS13]. It is shown that the auction is efficient under the assumption of price taking buyers. Furthermore, when the agents' utilities are strictly concave functions, the seminal study in [JT04] establishes a strong theoretical upper limit on the auction's loss of efficiency at ¼ of the maximum attainable social welfare.

This project was funded by NSF-CPS Award No. CNS-1136040 to Pahwa (PI), Das, Deloach, Natarajan, Ou. All correspondence should be addressed to sdas@ksu.edu.

More recently, it has been shown that even when buyers are price anticipating, proportional allocation allows the mechanism to attain the best possible outcome [JT09]. Similar theoretical limits have been investigated for a more general class of auctions called smooth auctions with proportional allocation [ST13]. Theoretical properties of a situation with multiple sellers participating in a proportional allocating auction, and with inelastic (fixed) demand for the resource has been examined [JT11].

When the proportional allocation auction takes into account the costs of the network's links through which the resource (data) flows, the efficiency is shown to be at least $4\sqrt{2}-5$ of the social welfare maximum with price anticipating buyers and with convex costs [JMT05]. In a separate study it has been shown that when the costs are linear, the mechanism's efficiency loss is lower bounded at $\frac{1}{3}$ of the maximum value [JT06]. Unfortunately, the previous studies are based on the assumption that the utility functions' are convex. Under a more general setting where this convexity assumption is not true, the mechanism's efficiency loss no longer enjoys a theoretical limit, and could in fact be arbitrarily large [CST15]. A few studies have proposed schemes to address an auction's efficiency loss arising from price anticipation. For instance an auction mechanism with price differentiation where each buyer has a different price, has been proposed [YML13]. This study also suggests a feedback control mechanism on the price vector that drives the auction to converge to the globally optimum social welfare.

Double auction mechanisms with proportional allocation have been studied [KV10]. However, this study considers only the case of price takers. The research reported in this paper includes a theoretical study on both price anticipating buyers and price anticipating sellers participating simultaneously in a double auction.

Research on energy auctions has closely followed the theoretical advancement in mechanism design. A large body of recent literature on energy auction algorithms model large utility company as the sellers [FAA14, JK15, Moh15, WLM15, ZX+15]. Many of these studies consider objectives and/or constraints that are applicable only to energy trade, such as generation scheduling [FAA14, Moh15], economic dispatch [ZX+15], and transmission losses [GM15]. Some energy auction studies are designed for revenue maximization and are not social welfare maximizing (efficient) mechanisms [JK15, ZX+15].

Efficient auctions have begun to gain research attention in the present context energy trade. Several such investigations do not consider proportionally fair allocation of energy [TS+12, TZ+14, WS+14]. One recent study proposes a VCG-style auction with multiple sellers and a single demand response aggregator as the buyer [NAC15]. Another study that uses the VCG mechanism reports a double auction [FK+14]. The cake-cutting algorithm has been applied to procure energy from multiple sellers and for a community of consumers acting in tandem as a single buyer [TY+15]. A truthful buyers' auction that makes use of the Arrow-d'Aspremont-Gerard-Varet mechanism has been suggested [MD+14].

While some existing approaches as well as this research explore Nash equilibrium, where all agents are assumed to act simultaneously [LCD15, NAC15, WH15], others use leader-follower games under Stackelberg equilibrium [TS+12, TZ+14, LG+15, WLM15]. This equilibrium concept is applicable when the energy market is modeled as an oligopoly with only a limited number of suppliers modeled as leader, or with the inclusion of an upper level agent such as an aggregator [TS+12, TZ+14, LG+15, NAC15]. This is in contrast to the present work that treats both buyers and sellers with equivalent market parity.

Many research papers in the existing research on energy auctions report only single-sided ones [MD+14, LCD15, Moh15, NAC15, TY+15, WH15, WLM15]. However, there are some papers that do address some form of double auction [DF+14, FK+14, GM15, JK15, LG+15, ZX+15]. One of them analyzes the loss of efficiency arising from Stackelberg equilibrium where the leading agent enjoys a first-mover's advantage in the underlying game [LG+15]. Another paper on double energy auctions, that proposes a primal-dual algorithm, is based on the unrealistic assumption that all agents – buyers and sellers, act in concert to lower the overall grid cost [15]. Yet another such study considers a double auction where the energy sellers are utility companies whose goal is revenue maximization [GM15]. A preliminary study on double energy auction based on the Kelly mechanism has been proposed [MF+14]. Another double auction study that does not use proportional allocation addresses software issues rather than the auction [DF+14].

Moreover, none of these single or double energy auctions address the issue of price anticipatory behavior of the participating agents. To the best of the authors' knowledge, only one research in the literature on efficient energy auctions does examine the adverse effect of price anticipation [LCD15]. This study is a single-sided auction with consumers of electricity acting as price anticipating buyers. The theoretical derivation of the auction's loss of efficiency closely follows that in earlier theoretical studies on single-sided auctions [JT04, JT11, CSS13].

A few of the extant research on energy auctions applies some form of price differentiation [FK+14, MF+14, TZ+14, GM15, TY+15]. Although this paper does not explicitly do so, it is worthwhile to mention that the research reported here can be readily extended to address this scenario as theoretical research on such mechanisms has been published [YML13].

*B. Contribution*

This research proposes a double auction mechanism that includes one set of agents as buyers, and another set as sellers. It also assumes the presence of a separate mediating agent called the *aggregator* whose role, unless indicated otherwise, is to (*i*) receive monetary bids from the buyers and available energy for trade from the sellers; (*ii*) proportionally allocate energy to the buyers; (*iii*) iteratively converge to the market clearing price.

The main contributions of this research are as follows.
(*i*) It performs a theoretical analysis of the equilibrium conditions arising from price anticipating buyers and sellers, thereby extending previous studies on single-sided auctions to double auctions. It shows the existence of a unique equilibrium for double auctions under price anticipation.
(*ii*) It proposes a distributed iterative auction algorithm, and suggests a possible realistic scheme through which the selfish agents may use information from prior iterations to approximate price anticipating behavior without any knowledge of the other agents.

(*iii*) It shows that, unlike in single sided auctions, double auctions can readily minimize the loss of efficiency arising from price anticipation, suggesting a simple extension of proportional allocation in order to approach the minimum.

(*iv*) It shows how the aggregator's own revenue can be incorporated within the auction framework, and proposes how this can lead to a bi-objective optimization problem where the aggregator is no longer strictly selfless by establishing the presence of a Pareto front within specific bounds.

It must be emphasized that although this research considers energy as the traded resource, the underlying theoretical analysis is directly applicable to other divisible resource auctions.

The rest of this paper is organized in the following manner. Section II presents the framework of the auction. The double auction under conditions of price anticipation and price taking are outlined in sections III and IV. The proofs of the propositions made in both sections have been postponed until the appendix. The results obtained from simulations are discussed in section V. Finally, the conclusion of this research is derived in section VI.

## II. FRAMEWORK

### A. Network Model

With energy as the resource involved in the trade, the network of agents in our model consists of a set $\mathcal{D}$ of buyers and another set $\mathcal{S}$ of sellers. Although grid energy auctions typically involve the presence of prosumers that buy and sell energy, we assume for simplicity that $\mathcal{D}$ and $\mathcal{S}$ are disjoint. The model also includes a separate entity, $\mathcal{A}$, the *aggregator* (or auctioneer) that is responsible for communicating with the other agents and implementing the auction. Unless otherwise indicated the aggregator acts as a selfless agent, requiring no separate parametrization of its own, in which case $\mathcal{A} = \emptyset$.

Each agent, whether a buyer or a seller, has its own utility representing the gain (in monetary units) it derives from consuming an amount of energy. The utility of a buyer $i \in \mathcal{D}$ is denoted as $u_i$, and that of a seller $j \in \mathcal{S}$, as $v_j$. As the sellers are capable of supplying energy, the model assumes that each has a fixed amount of energy $g_j$, called its generation that is available both for its own use and to sell.

The underlying physical network that implements the auction mechanism can be completely defined as the following 6-tuple $\Theta$,

$$\Theta \triangleq (\mathcal{D}, \mathcal{S}, g_j, u_i, v_j, \mathcal{A}). \quad (1)$$

The mathematical treatment made throughout the rest of this paper is based on the following underlying assumptions.

(*i*) The utilities $u_i$ and $v_j$ are continuous, differentiable, monotonically increasing and strictly concave functions with non-negative arguments. In other words, $u_i', v_j' > 0$ and $u_i'', v_j'' < 0$ when the argument lies within the interval $(0, \infty)$.

(*ii*) There is at least one buyer and one seller, i.e. $\mathcal{D}, \mathcal{S} \neq \emptyset$, and furthermore that at least one buyer $i \in \mathcal{D}$ that can obtain energy from some seller $j \in \mathcal{S}$ so that there some trade takes place. This assumption can be summarized as follows.

$$\exists i \in \mathcal{D}, j \in \mathcal{S}, \ni u_i'(0) > v_j'(g_j). \quad (2)$$

### B. Auction Process

The buyers' and sellers' bidding processes are implemented as separate steps in the auction. Each buyer $i$ receives from the aggregator its *demand* $d_i$, which is the amount of energy that it is allocated for use. The buyer responds by communicating to the aggregator its *bid* $b_i$, which is the amount of money that it is willing to pay for it. Separately, each seller $j$ receives a per unit *price* $p$ of energy, and communicates back to the aggregator, its *availability* $a_j$ that it is willing to supply.

The schematic below (Fig. 1.) shows the layout of the entire auction process. The auction proceeds iteratively until termination when $p$ has converged to the market clearing price.

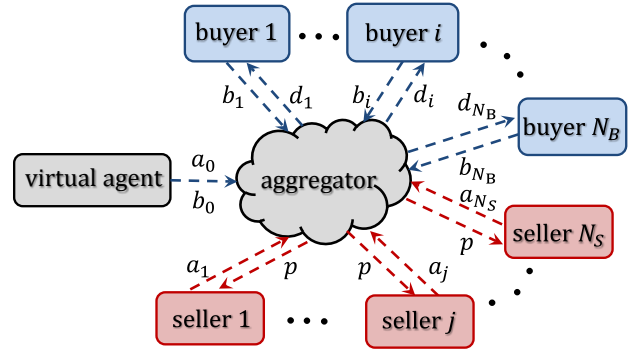

Fig.1: Schematic of the network model showing flow of information during auction.

## III. DOUBLE AUCTION UNDER PRICE ANTICIPATION

### A. Aggregator

It is assumed that there is no energy loss taking place during transmission. Thus, with the network operating under isolation as is also assumed in this section, the total amount of energy that is declared available by the sellers must be equal to the total amount demanded by the buyers, so that the following *energy balance* equation holds.

$$\sum_i d_i = \sum_j a_j. \quad (3)$$

In this section the aggregator is also assumed to be selfless and plays no additional role other than that specified earlier ($\mathcal{A} = \emptyset$), so that the money received as the total buyers' bids is exchanged for the total available energy sold by the sellers. Under these circumstances, the per unit price is given by,

$$p = \left(\sum_j a_j\right)^{-1} \sum_i b_i. \quad (4)$$

As the auction is based on proportional allocation of resources, the energy demand $d_i$ that each buyer $i$ receives from the aggregator must be proportional to its bid $b_i$ so that,

$$d_i = \frac{b_i}{p}, \quad \forall\, i \in \mathcal{D}. \quad (5)$$

### B. Buyer

Each buyer $i$ aims to maximize its payoff from the auction mechanism. Noting that it has to pay an amount $b_i$ in order to receive energy $d_i$, it places a bid $b_i$ in accordance with the following optimization problem.

Maximize w.r.t. $b_i$:
$$u_i(d_i) - b_i. \tag{6}$$

Proposition-1: The optimal bidding strategy of a buyer $i \in \mathcal{D}$ is,
$$b_i = d_i u_i'(d_i)(1 - \beta_i). \tag{7}$$

Here the quantity $\beta_i$ is the market power of buyer $i$ described later in this section.

### C. Seller

Each seller $j$ declares its availability $a_j$ at price $p$ to the aggregator to attain the maximum of its payoff, which is the sum of the money that it receives from selling energy as well as its own utility from consuming the remaining amount $g_j - a_j$ of energy. Noting that its availability cannot exceed its generation $g_j$, its participation in the auction is characterized by means of the following optimization problem.

Maximize w.r.t. $a_j$:
$$v_j(g_j - a_j) + p a_j. \tag{8}$$

Subject to:
$$a_j \leq g_j. \tag{9}$$

Proposition-2: The optimal bidding strategy of a seller $j \in \mathcal{S}$ is given by the expression below.
$$a_j = \min\{a_j^\circ, g_j\}, \tag{10}$$

where $a_j^\circ$ is the solution to the equation,
$$v_j'(g_j - a_j^\circ) = p(1 - \alpha_j). \tag{11}$$

The seller's market power $\alpha_j$ is described below.

### D. Market Power

The market power of an agent reflects its ability to influence the overall outcome of the auction. When the auction involves a large number of agents, an individual agent's action cannot exert a great deal of influence on the outcome; consequently the agent's market power is low. In the limiting case when there are an infinite number of agents, the market power approaches zero. It is this limiting case that price taking conditions serves to approximate.

In the present case, the market power $\beta_i$ of every buyer $i$ and $\alpha_j$ that of sellers can be defined through separate expressions, given below,

$$\beta_i = \left(\sum_{i'} b_{i'}\right)^{-1} b_i, \quad \forall i \in \mathcal{D}, \tag{12}$$

$$\alpha_j = \left(\sum_{j'} a_{j'}\right)^{-1} a_j, \quad \forall j \in \mathcal{S}. \tag{13}$$

The level of awareness of each buyer or seller about the remaining agents can vary from complete unawareness (price taking) to full awareness of the others' bidding strategies (as required in Eqns. (12) and (13) above). A realistic scenario lies somewhere in between. In such a case, the iterative auction would allow the buyer or seller to approximate its market power from the information gleaned from previous iterations. The expressions below, which are derived from Eqns. (7) – (11) can be used as the means by which each buyer or seller can obtain such estimates. Superscripts $(k-1)$ and $(k)$ have been introduced for clarity to indicate each iteration $k$ and its immediately preceding iteration $k-1$.

$$\beta_i^{(k)} = 1 - \frac{b_i^{(k-1)}}{d_i^{(k-1)} u_i'(d_i^{(k-1)})}, \tag{14}$$

$$\alpha_j^{(k)} = 1 - \frac{1}{p^{(k-1)}}\left(v_j'(g_j - a_j^{(k-1)}) - \rho_j^{(k-1)}\right). \tag{15}$$

Note that neither expression above incorporates quantities pertaining to the other agents present in the network.

The quantity $\rho_j$ in Eqn. (15) is a dual variable obtained from the constrained optimization problem in Eqns. (8) and (9). Further details pertaining to $\rho_j$ can be found in the appendix. It suffices to mention that $\rho_j = 0$ except in the case when the seller declares its entire generation as the availability, i.e. $a_j = g_j$.

At the onset of the auction process ($k = 1$), when the agents lack prior information, the market powers may be initialized to zero so that the agents act as simple price takers.

### E. Distributed Double Auction Algorithm

Before the iterative bidding process takes place, there are several ways by which the aggregator can initialize the auction variables $p$ and the $d_i$s that it communicates to the sellers and buyers so that they can place their bids. An effective way to minimize the number of steps would be to use stored historical information from previous rounds. Otherwise the aggregator may use heuristic means to do so. In the most simplistic case, these variables may be assigned entirely randomly. This initialization and the subsequent auction steps are outlined below.

---

**Distributed Double Auction Algorithm**

```
Initialize p^(0), d_i^(0) ∀ i ∈ D
//Buyer i ∈ D:   β_i ← 0
//Seller j ∈ S:  α_j ← 0
Set k ← 1
While (termination criterion = 'F')
    Send p^(k) to sellers j ∈ S
    //Seller j ∈ S bid
    Receive a_j^(k) from sellers j ∈ S
    Send d_i^(k) to buyers i ∈ D
    //Buyers i ∈ D bid
    Receive b_i^(k) from buyers i ∈ D
    Increment k ← k + 1
    Obtain d_i^(k)
    Update p^(k)
    //Buyers estimate β_i^(k)
    //Sellers estimate α_j^(k)
    Evaluate termination criterion
End
```

---

### F. Equilibrium

The auction steps described earlier terminates when further updates of neither the price $p$ nor any of the bids submitted by the agents to the aggregator are changed. This is when generalized Nash equilibrium [FK10] is established.

In order to characterize the equilibrium conditions under price anticipation, the functions $\pi_i(\cdot)$ and $\pi_j(\cdot)$ are introduced below,

$$\pi_i(d_i) = \left(1 - \left(\sum_j a_j\right)^{-1} d_i\right) u_i(d_i) + \left(\sum_j a_j\right)^{-1} \int_0^{d_i} u_i(z) dz, \quad (16)$$

$$\pi_j(g_j - a_j) = v_j(g_j - a_j) \left(\sum_{j'} a_{j'}\right) \left(\sum_{j' \neq j} a_{j'}\right)^{-1} - \left(\sum_{j' \neq j} a_{j'}\right)^{-1} \int_0^{a_j} v_j(g_j - z) dz. \quad (17)$$

The effect of price anticipation can be examined in terms of the following constrained optimization problem.

Maximize w.r.t. $d_i, a_j$:
$$\Pi(d_i, a_j | \Theta) = \sum_i \pi_i(d_i) + \sum_j \pi_j(g_j - a_j). \quad (18)$$

Subject to constraints in Eqns. (3) and (9) which are restated below,

$$\sum_i d_i = \sum_j a_j,$$
$$a_j \leq g_j.$$

Denoting the solutions of the above maximization problem as $d_i^\dagger$ and $a_j^\dagger$, the social welfare is $U^\dagger \triangleq U(d_i^\dagger, a_j^\dagger | \Theta)$. Additionally, the maximum social welfare corresponding to the efficient solution is denoted as $U^*$.

Proposition-3: (*i*) There exists a unique equilibrium of the double auction where the demand $d_i$ of each buyer $i \in \mathcal{D}$ and availability $a_j$ of each seller $j \in \mathcal{S}$ is the solution to the optimization problem defined in Eqn. (18) with Eqns. (3) and (9) as constraints.

(*ii*) The social welfare attained under price anticipation is no greater than that attainable under price taking, i.e.,
$$U^\dagger \leq U^*. \quad (19)$$
The above statement implies that there is a loss of efficiency due to price anticipation.

## IV. DOUBLE AUCTION UNDER PRICE TAKING

### A. Efficient Solution

The efficient solution can be obtained from the following constrained optimization problem.

Maximize w.r.t. $d_i, a_j$:
$$U(d_i, a_j | \Theta) = \sum_i u_i(d_i) + \sum_j v_j(g_j - a_j). \quad (20)$$

Subject to constraints in Eqns. (3) and (9) which are restated below.

$$\sum_i d_i = \sum_j a_j.$$
$$a_j \leq g_j, \quad \forall j \in \mathcal{S}.$$

For the sake of convenience the buyer $i$'s bidding strategy, which is that in Eqn. (7) with $\beta_i = 0$, is provided below.
$$b_i = d_i u_i'(d_i). \quad (21)$$
The seller $j$'s strategy is determined according to Eqns. (10) and (11) where $\alpha_j = 0$, and given below,

$$a_j = \min\{a_j^\circ, g_j\}, \quad (22)$$
where $a_j^\circ$ is the solution to the equation,
$$v_j'(g_j - a_j^\circ) = p. \quad (23)$$

Proposition-4: Under the assumption that the buyers and sellers are price takers, the following statements are true for the double auction.

(*i*) The buyer and seller strategies are defined according to Eqns. (21), (22) and (23).

(*ii*) The equilibrium demand $d_i^*$ of each buyer $i$ and availability $a_j^*$ of each seller $j$ after the termination of the auction are unique solutions of Eqns. (3), (9) and (20).

(*iii*) There is no loss in efficiency, i.e.
$$U^* \triangleq U(d_i^*, a_j^* | \Theta) = \max_{d_i, a_j} U(d_i, a_j | \Theta). \quad (24)$$

Thus the unique equilibrium of the double auction is also the efficient solution.

The market price at equilibrium from the auction is denoted as $p^*$. At equilibrium, the derivative of the utility function (also called *marginal utility*) of each buyer and that of each seller that is not trading its entire generation $g_j$ is equal to the market price; and for traders that trade all of it, more than the price. Mathematically,

$$\begin{cases} u_i'(d_i^*) = p^*; \\ v_j'(g_j - a_j^*) = p^* \text{ when } a_j^* < g_j; \\ v_j'(g_j - a_j^*) > p^* \text{ when } a_j^* = g_j; \end{cases} \quad (25)$$

The equilibrium can be understood readily graphically as shown in Fig. 2(*a*). Here, we define the aggregate *demand function* $D(p)$ as the total amount of energy delivered to the buyers as a function of the market price $p$. Likewise, we define the *availability function* $A(p)$ as the total availability declared by the suppliers as a function of $p$. Thus,

$$D(p) = \sum_i d_i, \quad (26)$$
$$A(p) = \sum_j a_j. \quad (27)$$

Proposition-5: Under the assumption of price taking, the following statements are true for the double auction.

(*i*) The availability function $A(p)$ is zero when $p \leq \min_j v_j'(g_j)$, monotonically increasing with price $p$ in the interval $p \in (\min_j v_j'(g_j), \max_j v_j'(0))$ and constant when $p \geq \max_j v_j'(0)$. In other words,

$$\begin{cases} A(p) = 0, & p \leq \min_j v_j'(g_j); \\ A(p) \text{ mon. inc.}, & \min_j v_j'(g_j) < p < \max_j v_j'(0); \\ A(p) \text{ constant}, & p \geq \max_j v_j'(0). \end{cases} \quad (28)$$

(*ii*) The demand function $D(p)$ is monotonically decreasing with price $p$ in the interval $p \in (0, \max_i u_i'(0))$ and zero when $p \geq \max_i u_i'(0)$.

$$\begin{cases} D(p) \text{ mon. dec.}, & p < \max_i u_i'(0), \\ D(p) = 0, & p \geq \max_i u_i'(0). \end{cases} \quad (29)$$

(*iii*) At the unique equilibrium price $p^*$, $A(p^*) = D(p^*)$.

### B. Virtual Bidding

In general, the loss of efficiency, when $d_i$ is the demand of each buyer $i$, and $a_j$, the allocation of each seller $j$, can be expressed as follows.

$$L_\Theta(d_i, a_j) = \frac{U(d_i^*, a_j^*|\Theta) - U(d_i, a_j|\Theta)}{U(d_i^*, a_j^*|\Theta)}. \tag{30}$$

The loss that takes place when the agents participate in the auction as price anticipators is $L_\Theta(d_i^\dagger, a_j^\dagger)$. This section shows how the basic proportional allocation double auction mechanism can be extended to mitigate the loss of efficiency.

In order to minimize the loss $L_\Theta$, a *virtual agent* can be introduced to the network defined earlier in Eqn. (1). The virtual agent, which is indexed with the subscript '0', participates in the auction simultaneously as a buyer and a seller with arbitrarily large availability $a_0$. As the virtual agent is incorporated within the aggregator, we let $\mathcal{A} = \{a_0\}$.

Since the virtual agent does not have its own generation, it buys back the amount of energy $a_0$ declared as its availability at the market price defined in Eqn. (4), so that,

$$b_0 = pa_0. \tag{31}$$

Proposition-6: As the virtual agent's availability $a_0$ increases, the loss in efficiency $L_\Theta$ from price anticipation decreases. In the limiting case,

$$\lim_{a_0 \to \infty} L_\Theta(d_i^\dagger, a_j^\dagger) = 0. \tag{32}$$

Since the inclusion of virtual bidding allows the auction to behave like a price taking mechanism, for the remainder of this section we assume that the buyers and sellers behave as price takers.

### C. Surcharge

We now consider the situation where the aggregator, $\mathcal{A}$, is no longer a strictly selfless enabler in the auction process but also has its own incentive to implement the mechanism by levying a *surcharge* price $p_s$ per unit of energy traded. Thus in the model in Eqn. (1) the aggregator now includes the surcharge, which we indicate by letting it be given by $\mathcal{A} = \{a_0 \to \infty, p_s\}$. The total revenue earned by the aggregator from the auction with the introduction of surcharge is given by,

$$R = p_s \sum_j a_j. \tag{33}$$

The expression for the price in Eqn. (4) is modified to account for the surcharge as follows,

$$\sum_i b_i = (p + p_s) \sum_j a_j. \tag{34}$$

With proportional allocation, the demand $d_i$ that each buyer $i$ receives is given by the following expression that replaces the earlier Eqn. (5),

$$d_i = \frac{b_i}{p_s + p}. \tag{35}$$

Fig. 2(b) illustrates the effect of the surcharge. Eqn. (35) shows that the buyers purchase energy at an effective per unit price of $p_s + p$ which is higher than $p$ that the sellers receive per unit of energy traded. The volume of energy traded is equal to $D(p_s + p) = A(p)$, which is lower than $A(p^*) = D(p^*)$.

We show that the price taking auction is the solution to the following constrained optimization problem.

Maximize w.r.t. $d_i, a_j$:

$$\Omega(d_i, a_j|\Theta) = \sum_i u_i(d_i) + \sum_j v_j(g_j - a_j) - p_s \sum_j a_j. \tag{36}$$

Subject to constraints in Eqns. (3) and (9),

$$\sum_i d_i = \sum_j a_j,$$
$$a_j \leq g_j.$$

Proposition-7: Under the assumption of price taking, the following statements are true with surcharge price $p_s > 0$.

(*i*) The buyer and seller strategies are defined according to Eqns. (21), (22) and (23).

(*ii*) The equilibrium demand $d_i$ of each buyer $i$ and availability $a_j$ of each seller $j$ of the auction are unique solutions of Eqns. (3), (9) and (36).

(*iii*) There exists a Pareto front where any increase in revenue $R$ is associated with a simultaneous decrease in the social welfare in Eqn. (20).

(*iv*) There exists an optimal surcharge $p_s^{\text{OPT}}$ that maximizes the aggregator $\mathcal{A}$'s revenue $R$.

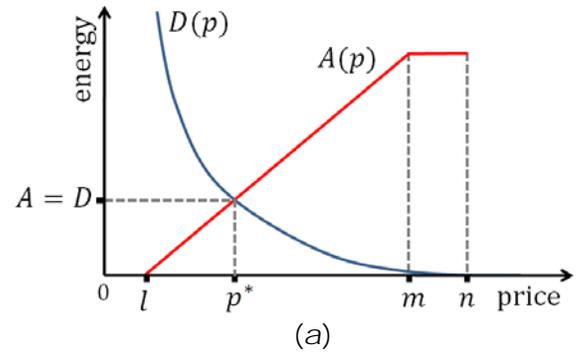

(a)

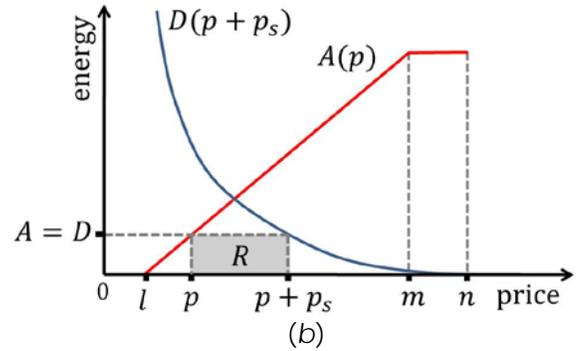

(b)

Fig.2: Plots of the $A(p)$ and $D(p)$ as functions of $p$. In the x-axis, $l = \min_j v_j'(g_j)$, $m = \max_j v_j'(0)$, and $n = \max_i u_i'(0)$. (*a*) Equilibrium conditions under price taking. (*b*) The addition of surcharge price with revenue $R$ being the area shaded in grey.

## V. RESULTS

### A. Setup

In order to compliment the theoretical considerations in the earlier sections, several sets of simulations were carried out. A total of five scenarios were considered, where the number of buyers and sellers were $|\mathcal{D}| = 2, |\mathcal{S}| = 3$, $|\mathcal{D}| = 2, |\mathcal{S}| = 6$, $|\mathcal{D}| = 2, |\mathcal{S}| = 10$, $|\mathcal{D}| = 3, |\mathcal{S}| = 2$ and $|\mathcal{D}| = 4, |\mathcal{S}| = 4$. In order to analyze the effect of price anticipation, the total number of agents were made relatively small in comparison to other simulation studies. Moreover, the first three scenarios contain only 2 sellers. This reflects the situation is a realistic microgrid, where the number of PV-equipped units is usually lower than the number of those without it. The fourth and fifth scenarios were added to explore the performance of the double auction under other potential situations.

The utilities of the buyers and sellers assumed to follow logarithmic saturation curves according to,
$$u_i(d_i) = x_i \log(y_i d_i + 1), \quad (42)$$
and,
$$v_j(g_j - a_j) = x_j \log(y_j(g_j - a_j) + 1). \quad (43)$$
The quantities $x_i, y_i, x_j$ and $y_j$ were different for each agent, and were generated randomly from a uniform distribution centered at unity. The generations, $g_j$, for the sellers were also drawn in at random, uniformly in the interval $[g_{min}, g_{max}]$.

### B. Price Anticipation

The first set of simulations was performed to examine the effect of price anticipation of the buyers and sellers upon the double auction. The results of this study are shown in Fig. 3. It can be seen that in each case there is a reduction in the social welfare due to price anticipation. More detailed analysis shows that when considered separately, while the social welfare of the buyers reduces due to price anticipation, the social welfare of the sellers is increased. This is because Propositions 1 and 2 indicate that price anticipation ($\beta_i, \alpha_j > 0$) causes the values of $d_i$ and $a_j$ to be lower than with price taking ($\beta_i, \alpha_j = 0$). Consequently, the volume of energy being traded is also less so that the surplus amount of energy $g_j - a_j$ remaining with each seller $j$ is higher, which also increases its utility, $v_j(g_j - a_j)$.

Although, due to price anticipation, the social welfare in a double auction is lower than its optimal value, the utilities of the sellers change in the opposite direction. Thus an observation made from this study is that the effect of price anticipatory agents in double-auctions is less severe in comparison to single-sided auctions.

### C. Virtual Bidding

The effect of virtual bidding was investigated through a second set of simulations. The results of these simulations are provided in Fig. 4, separately for each of the five scenarios.

It can be seen that the loss in efficiency $L_\Theta$ approaches zero as $a_0$ increases towards $a_0 \to \infty$. This observation holds true for each of the five scenarios that were simulated, and is consistent with Proposition-6.

### D. Surcharge

In order to examine the role of the surcharge price $p_s$ on the double auction, a set of simulations were carried out for each of the five scenarios described earlier. As price taking conditions are assumed, the agents' market powers were always set at $\beta_i = 0, \alpha_j = 0$, throughout the iterative mechanism.

The auction was simulated until equilibrium for different values of the surcharge price $p_s$. Fig. 5 shows the Pareto front discussed in the claim (*iii*) in Proposition-7. In each scenario, the extreme left end points of the fronts correspond to $p_s = 0$ so that the social welfare is maximum $U = U^*$, while the aggregator's revenue $R = 0$.

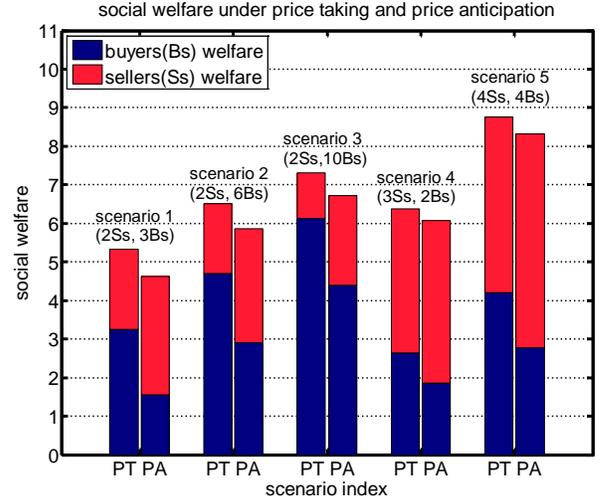

Fig.3: Social welfare $U$ under price taking (PT) and price anticipation (PA) for each scenario.

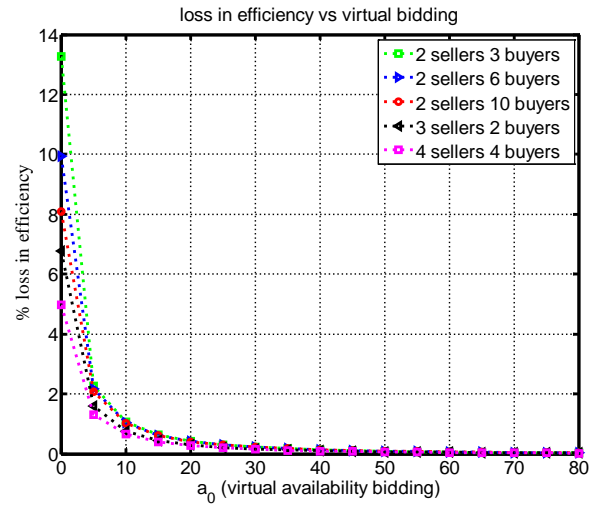

Fig.4: Loss in efficiency $L_\Theta$ as a function of $a_0$ for each scenario.

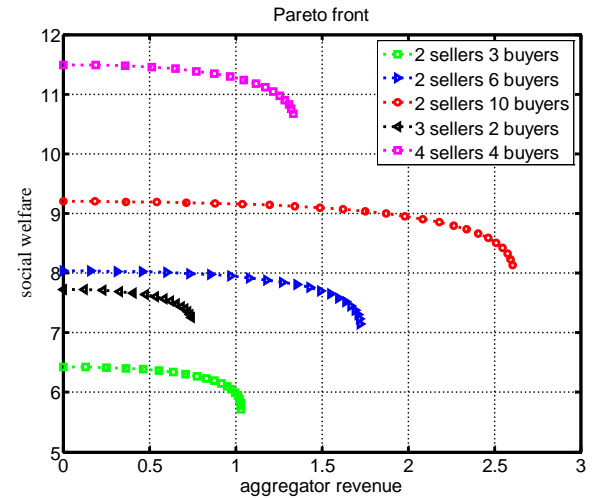

Fig.5: The Pareto front of the revenue $R$ and social welfare $U$ with varying surcharge price $p_s$, for each scenario.

As $p_s$ progressively increases until $p_s^{OPT}$, so does $R$, while $U$ decreases. The right ends of the Pareto fronts correspond to $p_s = p_s^{OPT}$. When $p_s$ exceeds $p_s^{OPT}$, both $U$ and $R$ decrease, which is not shown in Fig. 5.

Fig. 6 shows how the aggregator's revenue $R$ varies with surcharge $p_s$. In each scenario, $R$ increases with $p_s$ until it reaches its maximum when the surcharge is $p_s^{OPT}$. In all but one scenario, the revenue $R$ can be seen to decrease beyond its corresponding maximum. These results are consistent with claim (*iv*) of Proposition-7.

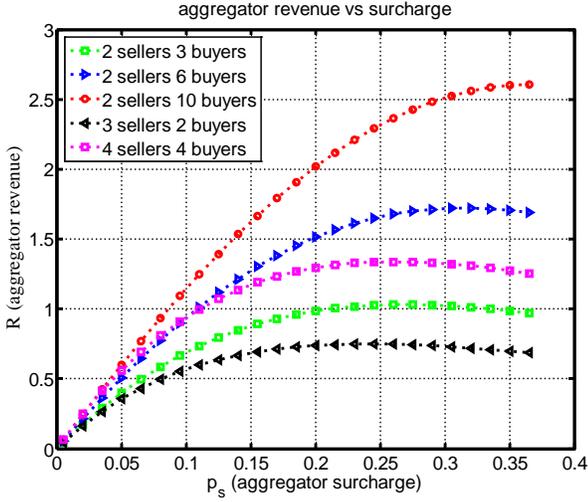

Fig.6: Aggregator's revenue $R$ as a function of surcharge price $p_s$, for each scenario.

## VI. CONCLUSION

The distributed double auction algorithm in Section-III can be implemented readily by the aggregator, even in the presence of a virtual agent or with surcharge pricing. The algorithm can optionally consider price-anticipatory agents. A possible method by which real world agents may use information gathered from earlier iteration to imitate price anticipation has been suggested.

It is shown that with price anticipating agents, the double auction's equilibrium coincides with that of a constrained optimization problem whose objective function $\Pi$ is different from the social welfare function $U$, resulting in a loss of efficiency. It is shown that when the aggregator incorporates a virtual agent that is simultaneously both a buyer and a seller, can minimize the loss of efficiency, so that the double auction can reach the efficient equilibrium.

A generalized auction scenario where the aggregator receives a surcharge price is investigated where in the limiting case, the aggregator may act as a selfish agent trying to maximize its revenue $R$ from the auction. With the social welfare $U$ and revenue $R$ as independent objectives, a bi-objective framework for the double auction mechanism is suggested.

The theoretical analysis has been supplemented by several simulations. The results of the simulations are in complete agreement with the theory.

## VII. APPENDIX

Let,
$$A = \sum_{j}^{N_B} a_j.$$
$$B = \sum_{i=1} b_i.$$

Proof of Proposition-1: The stationary condition of Eqn. (6) is obtained by differentiation with respect to the bid $b_i$ as shown below.
$$u_i'(d_i)\frac{\partial d_i}{\partial b_i} = 1.$$
As buyer $i$ is price anticipating, $d_i$ is dependent on $b_i$ through the price $p$. Hence replacing $\frac{\partial d_i}{\partial b_i}$ appropriately using Eqn. (5) and applying the chain rule we get,
$$u_i'(d_i)\frac{1}{p}\left(1 - \frac{b_i}{p}\frac{\partial p}{\partial b_i}\right) = 1. \tag{A1}$$
Using Eqn. (4) the above equality yields,
$$u_i'(d_i)\left(1 - \frac{b_i}{B}\right) = p. \tag{A2}$$
Whence from Eqn. (12),
$$u_i'(d_i)(1 - \beta_i) = p. \tag{A3}$$
Proposition-1 follows directly from the above and Eqn. (5). ∎

Proof of Proposition-2: Introducing the dual variable $\rho_j$, the Lagrangian of the problem defined in Eqns. (8) and (9) is,
$$\mathcal{L}_S(a_j, \rho_j) = v_j(g_j - a_j) + pa_j + \rho_j(a_j - g_j).$$
This yields the following KKT conditions.
$$\rho_j(a_j - g_j) = 0, \tag{A4}$$
$$v_j'(g_j - a_j) = p + a_j\frac{\partial p}{\partial a_j} + \rho_j. \tag{A5}$$
Replacing $\frac{\partial p}{\partial a_j}$ above appropriately using Eqn. (4),
$$v_j'(g_j - a_j) = p(1 - \alpha_j) + \rho_j. \tag{A6}$$
The quantity $\alpha_j$ is defined in Eqn. (13). When $a_j < g_j$, Eqn. (A4) shows that $\rho_j = 0$. The corresponding availability in Eqn. (A6) is equal to $a_j^\circ$ that solves Eqn. (11). The other situation in (A4) arises when $\rho_j < 0$, in which case the entire generated energy is declared available, i.e. $a_j = g_j$.

We can rewrite the above observations more concisely as,
$$\begin{cases} \rho_j = 0, & a_j < g_j; \\ \rho_j < 0, & a_j = g_j. \end{cases} \tag{A7}$$
∎

Proof of Proposition-3: Observe that, using Eqns. (3), (12) and (13), the derivatives of the functions defined earlier in Eqns. (16) and (17) are,
$$\frac{\partial}{\partial d_i}\pi_i = (1 - \beta_i)u_i'(d_i), \tag{A8}$$
$$\frac{\partial}{\partial a_j}\pi_j = -\frac{1}{1 - \alpha_j}v_j'(g_j - a_j). \tag{A9}$$
From Eqn. (12) since $\beta_i > 0$, whenever $d_i > 0$, $\frac{\partial}{\partial d_i}\pi_i > 0$ in Eqn. (A8). The factor $(1 - \beta_i)$ in Eqn. (A8) is also strictly decreasing in $b_i$ and hence $d_i$. Thus $\frac{\partial}{\partial d_i}\pi_i$ is also monotonically decreasing. Therefore $\pi_i$ is a strictly concave function. In a similar manner, from Eqn. (13) it is clear that $\alpha_j < 1$ as long as $0 \leq a_j \leq g_j$, so that $\frac{\partial}{\partial a_j}\pi_j < 0$ in Eqn. (A9). Besides as $\frac{1}{1-\alpha_j}$ is strictly increasing, the product is monotonically decreasing. Therefore $\pi_j$ is strictly concave. Thus there is a unique maximum of $\Pi(d_i, a_j|\Theta)$ as defined in Eqn. (18).

The Lagrangian of the problem defined in Eqn. (18), with Eqns. (3) and (9) acting as constraints, is given by,

$$\mathcal{L}(d_i, a_j, \lambda_j, \mu) = \sum_i \pi_i(d_i) + \sum_j \pi_j(g_j - a_j)$$
$$+ \sum_j \lambda_j(a_j - g_j)$$
$$+ \mu\left(\sum_j a_j - \sum_i d_i\right). \quad (A10)$$

The quantities $\mu$ and $\lambda_j$ above are the dual variables introduced by the constraints in Eqns. (3) and (9). The primal conditions from Eqns. (3) and (9) must be satisfied. Furthermore, complementary slackness conditions yield,

$$\lambda_j(a_j - g_j) = 0. \quad (A11)$$

From Eqns. (A8) and (A9), the stationary conditions of Eqn. (A10) must satisfy,

$$(1 - \beta_i)u_i'(d_i) = \mu, \quad (A12)$$
$$v_j'(g_j - a_j) = (1 - \alpha_j)(\lambda_j + \mu). \quad (A13)$$

From Eqns. (A3) and (A12) it is observed that $\mu = p$. Using Eqn. (5) it is seen that the buyer's bidding strategy defined in Eqn. (7) is satisfied.

When $a_j < g_j$, Eqn. (A11) shows that $\lambda_j = 0$. Replacing $\lambda_j$ and $\mu$ with 0 and $p$, Eqn. (11) is satisfied. On the other hand, when $a_j = g_j$, $\lambda_j$ is set to an appropriate value. From the concavity assumption, $v_j'(g_j - a_j) < p(1 - \alpha_j)$ so that $\lambda_j < 0$. We summarize these observations as follows.

$$\begin{cases} p = \mu; \\ \lambda_j < 0 \text{ when } a_j = g_j; \\ \lambda_j = 0 \text{ when } a_j < g_j. \end{cases} \quad (A14)$$

From the above considerations, it is seen that in both cases Eqn. (A13) satisfies the seller's bidding strategy in Eqn. (10). Eqn. (19) is trivially true since $U^*$ which is defined in Eqn. (24) is the maximum social welfare. ∎

Proof of Proposition-4: First, note that since the utilities are strictly concave, there is a unique optimum of the optimization problem defined in Eqn. (20) with constraints defined in Eqns. (3) and (9). The Lagrangian is,

$$\mathcal{L}(d_i, a_j, \lambda_j, \mu) = \sum_i u_i(d_i) + \sum_j v_j(g_j - a_j)$$
$$+ \sum_j \lambda_j(a_j - g_j)$$
$$+ \mu\left(\sum_j a_j - \sum_i d_i\right). \quad (A15)$$

The stationary conditions satisfy,

$$\lambda_j(a_j - g_j) = 0, \quad (A16)$$
$$u_i'(d_i) = \mu, \quad (A17)$$
$$v_j'(g_j - a_j) = \lambda_j + \mu. \quad (A18)$$

Comparing Eqn. (A17) with Eqn. (21), under proportional allocation in Eqn. (5), we see that $\mu = p$. Replacing $\frac{\partial p}{\partial a_j}$ in Eqn. (A5) with zero, the seller's stationary conditions satisfy (A4) and the following,

$$v_j'(g_j - a_j) = p + \rho_j. \quad (A19)$$

From Eqns. (A17), (A18) and (A19),

$$\begin{cases} p = \mu; \\ \lambda_j < 0 \text{ when } a_j = g_j; \\ \lambda_j = 0 \text{ when } a_j < g_j. \end{cases} \quad (A20)$$

Statements (*i*) and (*ii*) follow from the above. Since the social welfare $U(d_i, a_j|\Theta)$ is maximized, statement (*iii*) holds. ∎

Proof of Proposition-5: For each seller $j$, from Eqn. (A7) it is seen that $\rho_j = 0$ when $a_j < g_j$. Also $\alpha_j = 0$ under price taking. Hence from Eqn. (A6),

$$a_j = g_j - v_j'^{-1}(p). \quad (A21)$$

Since $v_j'' > 0$, $a_j$ is strictly increasing in the interval $p \in \left(v_j'(g_j), v_j'(0)\right)$. Moreover as $a_j \in [0, g_j]$ and as Eqn. (26) shows, $A(p)$ is the sum of all $a_j$s, statement (*i*) follows.

For each buyer $i$, from Eqn. (A3), with $\beta_i = 0$,

$$d_i = u_i'^{-1}(p). \quad (A22)$$

Since $u_i'' > 0$, $d_i$ is strictly decreasing with $p$ in the interval $p \in (0, u_i'(0))$. Moreover $d_i = 0$ when $p \geq u_i'(0)$. Hence statement (*ii*) follows directly from Eqn. (23) where $D(p)$ is expressed as the sum of all $d_i$s.

From Eqn. (2), there is a non-empty interval $p \in \left(\min_j v_j'(g_j), \max(\max_i u_i'(0), \max_j v_j'(0))\right)$ within which $D(p)$ is monotonically decreasing or zero and $A(p)$ is monotonically increasing or fixed at a positive value. Thus there is a unique $p^*$ such that $A(p^*) = D(p^*)$. ∎

Proof of Proposition-6: Since from Eqn. (31) $b_0 = pa_0$, the expression for the price in Eqn. (4) is replaced with,

$$p = \left(a_0 + \sum_j a_j\right)^{-1}\left(b_0 + \sum_i b_i\right). \quad (A23)$$

With the addition of virtual bidding, Eqns. (12) and (13) pertaining to market powers are rewritten as,

$$\beta_i = \left(b_0 + \sum_{i'} b_{i'}\right)^{-1} b_i, \quad \forall i \in \mathcal{D}, \quad (A24)$$

$$\alpha_j = \left(a_0 + \sum_{j'} a_{j'}\right)^{-1} a_j, \quad \forall j \in \mathcal{S}. \quad (A25)$$

Eqns. (A24) and (A25) show that both $\beta_i$ and $\alpha_j$ monotonically decrease with increasing $a_0$. From Eqns. (A8) and (A9) it follows that $\frac{\partial}{\partial d_i}\pi_i$ and $\frac{\partial}{\partial a_j}\pi_j$ monotonically approach $u_i'$ and $v_j'$. As we have shown that $\pi_i$ and $\pi_j$ are strictly concave, it follows that they increase monotonically with increasing values of $a_0$; whence from Eqn. (18), $U^\dagger$ also increases monotonically. In the limiting case, $\lim_{b_0 \to \infty} \beta_i = 0$, and $\lim_{a_0 \to \infty} \alpha_j = 0$; whereupon it follows that

$\lim_{a_0 \to \infty} U^\dagger = U^*$. Simultaneously, from Eqn. (30), the loss of efficiency decreases monotonically towards zero. This shows that inserting the virtual bidder into the auction allows the auction to simulate price-taking.

∎

Proof of Proposition-7: The equilibrium of the optimization problem defined in Eqn. (36) is unique because the addition of the linear term involving $p_s$ does not alter the concavity property.

The Lagrangian of the problem defined in Eqn. (36), with Eqns. (3) and (9) as constraints, is given by,

$$\mathcal{L}(d_i, a_j, \lambda_j, \mu) = \sum_i u_i(d_i) + \sum_j v_j(g_j - a_j) - p_s \sum_j a_j \\ + \sum_j \lambda_j(a_j - g_j) \\ + \mu \left( \sum_j a_j - \sum_i d_i \right). \quad \text{(A26)}$$

The stationary conditions satisfy,

$$\lambda_j(a_j - g_j) = 0, \quad \text{(A27)}$$
$$u_i'(d_i) = \mu + p_s, \quad \text{(A28)}$$
$$v_j'(g_j - a_j) = \lambda_j + \mu. \quad \text{(A29)}$$

Analogous to the reasoning provided in the proof of Proposition-4, it can be established that at equilibrium we must have,

$$\begin{cases} p = \mu; \\ \lambda_j < 0 \text{ when } a_j = g_j; \\ \lambda_j = 0 \text{ when } a_j < g_j. \end{cases} \quad \text{(A30)}$$

This establishes the statements (*i*) and (*ii*).

From Eqn. (33), when $p_s = 0$, the revenue $R = 0$. However, from the assumption in Eqn. (2), the aggregate availability $A$ is nonzero. Thus the social welfare is at the unique maximum $U^* > 0$. Increasing $p_s$ monotonically increases the revenue $R$ and also monotonically decreases the social welfare $U$. This shows the existence of a non-singleton Pareto front as claimed in (*iii*).

It can be readily inferred from Fig. 2(*b*) that for a sufficiently large value of $p_s$, the aggregate demand is zero, so that the volume of energy traded is zero and $R = 0$. In fact the upper limit of $p_s$ is defined as,

$$p_s < \max_i u_i'(0) - \min_j v_j'(g_j). \quad \text{(A31)}$$

It is concluded that there is an optimal $p_s$ that maximizes the aggregator's revenue $R$ verifying the claim in statement (*iv*).

∎